\documentclass{ws-procs975x65}

\begin{document}

\title{QUANTA: THE ORIGINALITY OF EINSTEIN'S APPROACH TO RELATIVITY?}

\author{CHRISTIAN  BRACCO $^*$}

\address{Art\'emis, CNRS, UNS, Observatoire de la C\^ote d'Azur, Bd de l'’Observatoire,\\
Nice,  06304, France\\
Syrte, CNRS, Observatoire de Paris, 61 av de l’'Observatoire,\\
Paris, 75014, France\\
$^*$E-mail: cbracco@unice.fr}

\begin{abstract}
\textit{To appear in the 13th Marcel Grossmann  Proceedings, Stockholm $1^\mathrm{st}-7^\mathrm{th}$ July} 2012 \\
We suggest that not only quanta may have played a role in Einstein's ideas on relativity, but that they themselves may be related to the dynamical and  relativistic behaviour of the electromagnetic field exhibited in a Poincar\'e's 1900 paper, in particular to the identical transformation law of energy and frequency for bounded plane waves. 
\end{abstract}

\bodymatter
\bigskip 

\par In the absence of any reliable document, from which Einstein'’s ideas on relativity before 1905 could uniquely be drawn, we are forced to make assumptions. If many historians (J. Norton \cite{Norton}, O. Darrigol \cite{OD}, \ldots) have priviledged an electromagnetic origin, others like A. Miller \cite{Miller} have emphasized the role of quanta. Certainly Einstein never mentioned quanta in his relativity papers, but as A. Pa\"is \cite{Pais} notes about his twofold (wave and energetic) derivation of the gravitational Doppler shift in 1911: ``[He] \textit{could not have totally forgotten it} [the relation $E=h\nu$ in the March 1905 paper] \textit{ - he never completely left the quanta hypothesis. But it was always in his style to avoid as much as possible to use this theory}''. However the presence of quanta may be comforted by three remarks  (see Ref.~\refcite{Bracco}). Firstly, the use of  the term ``\textit{light complex}"(bounded plane wave) when Einstein calculated the ``\textit{Transformation of the energy of Light Rays}", title of the section 8 of his June 1905 relativity paper \cite{Einstein}; then he said in passing that ``\textit{It is noteworthy that the energy and the frequency of a light complex vary with the observer state of motion according to the same law}", which for us today means the relativistic covariance of the quanta hypothesis. Miller calls it ``\textit{one of the great understatements in the history of science}" and indeed  in his 1909 Salzburg conference Einstein acknowledged that conceiving ``\textit{the radiation as very few extended complexes of energy $h\nu$ \ldots allows to have a very rough but concrete idea of the quanta hypothesis}". Secondly, the second postulate according to which ``\textit{Every light ray moves in the ``rest" coordinate system with a fixed velocity $c$, independently of whether this light ray is emitted by a body at rest or in motion \ldots}"\cite{Einstein} endowes by fact light quanta with a precise velocity $c$ (that of waves), a point left apart by Einstein in March \cite{Einstein}. Finally, the  ether is not rejected as a ``\textit{superflous}" hypothesis by him solely in regard to the absence of an absolute frame: freely moving quanta do not require it; reciprocally when he doubted of them in a well-known letter to Hopf in 1909, he concluded that the ether may have ``\textit{hope for a new life}". These remarks strongly suggest that quanta and relativity worked together, in a kind of fruitful dialectic in Einstein'’s mind. In the preface of a book by J. Stachel \cite{Stachel} on Einstein''s 1905 papers, R. Penrose already emphasized that Einstein ``\textit{appears to have had very clear and profound ideas as to what Nature was ``really like" at levels not readily perceivable by other physicists} \ldots  \textit{To me, it is virtually inconceivable that he would have put forward two papers} [March/June] \textit{in the same year which depended upon hypothetical views} [particle/wave] \textit{of Nature that he felt were in contradiction with each other}". 
\par In this paper, I defend the idea that the association of quanta and relativity is natural if one has in mind Planck'’s work on the black body radiation and Poincar\'e'’s contribution \cite{Poincare} \textit{Lorentz theory and the reaction principle} to the Lorentz Jubilee, which both occurred in mid-December 1900. Let us recall that whereas Planck introduced fixed energy quanta $\epsilon= h\nu$ from the energy dependence of the black body entropy (in a statistical perspective), Einstein came back in March 1905  to Wien'’s 1896 law in order to focalise on the (gas-like) volume dependence of this entropy \cite{Einstein}. Looking for moving quanta in such a way is of course part of an atomic and thermodynamic view of Einstein on Nature, but it may have been encouraged by the fact that Poincar\'e's 1900 paper not only opened as well known the way to a full dynamical and relativistic account of the electromagnetic field behaviour, but also exhibited  at first order in $v/c$ \textbf{the same transformation law for the energy and the frequency} of an electromagnetic plane wave of finite length. 

\par In this paper, analysed by O. Darrigol\cite{OD}, Poincar\'e aimed at solving the paradox that Lorentz theory did not satisfy the principle of reaction, which was a major failure in his eyes since it would allow perpetual motion. He first showed from Maxwell equations that the conservation of momentum for matter and electromagnetic field is satisfied if the latter is endowed with the momentum density $E \wedge B$ (we put below $c=1$). But since in classical mechanics the principle of reaction is linked to that of relativity (and to energy conservation), he came back to Lorentz 1895 fictitious ``\textit{local time}"  $t'’=t-vx$ which he interpreted as that indicated by the clocks of two distant observers synchronizing them with light signals, but ignoring their translational motion. (This amounts correcting Galilean relativity so that the velocity $c$ is preserved at first order in $v$ in any inertial frame). At the end of the paper, Poincar\'e analysed the emission of a plane wave of ``\textit{real}"  length $L$ by an Hertzian oscillator at the focus of a parabolic mirror (the ``\textit{gun}") moving at velocity $v'$ in the rest (ether) frame. He showed, by using the local time $t'$ and the transformation laws for the fields, that in a frame moving at velocity $v$, the ``\textit{apparent}" length $L'$ and the ``\textit{apparent}" energy $E'$  ($J'\Delta t$ in his notations) of the perturbation are related to the real ones by 
\begin{equation}
L’'= L(1+v) \hspace{1cm}  ;  \hspace{1cm}E ’ '= E(1-v).
\label{cb:eq1}
\end{equation}
Miller emphasized that Poincar\'e ``\textit{was the first to deduce the Lorentz transformation of a light pulse}" but he did not notice that, as J.-P. Provost first pointed out to me in a private communication, $E$ transforms like frequency or like $L^{-1}$ (usual Doppler effect) in Poincar\'e's paper. It must be said in Miller's defense that Poincar\'e did not comment Eq. (\ref{cb:eq1}) because he was interested by the compatibility of the momentum conservation laws in the two frames which he wrote $m\Delta v'= -E$ and $m\Delta (v'-v)=-E(1-v)$. He justified it by invoking  a ``\textit{complementary force}"  (the Li\'enard force coming from the application of local time to the dynamic of charges), instead of considering the true momentum changes $\Delta (mv')$ and $\Delta [m(v'-v)]$ and introducing as today a mass variation of the gun $\Delta m =-E$. In a paper of May 1906 (analysed in Ref.~\refcite{OD}), Einstein mentioned for the first time Poincar\'e's 1900 paper considering that ``\textit{the simple and formal considerations} [developed here] \textit{are already included in substance in a work of Henri Poincar\'e}"\cite{Einstein}.  Like Poincar\'e, he compared the conservation law of momentum for a system of charges and radiation with the law of the uniform motion of its center of mass (radiation being provided with a mass density $(E^2+B^2)/2)$. Whereas Poincar\'e considered that the Joule contribution $\int xj.E d\tau$ invalidated the latter, Einstein recovered it by considering that the Joule effect contributed to the internal energy of the particles i.e. to their masses, and consequently by writing this contribution as $\Sigma x_idm_i/dt$. For this interpretation, he refered to his September paper where he considered the emission of two opposite ``\textit{systems of plane waves}" by a body at rest (a symmetric extension of Poincar\'e's gun) and were he was led to claim that ``\textit{if a body emits the energy $L$ in the form of radiation, its mass decreases by $L$}". 
\par In guise of conclusion, it can be said from a scientific perspective that March, June and September 1905 Einstein's papers constitute a remarkable synthesis, in a fully original and new epistemological view, of Planck's 1900 work on the interaction between matter and radiation in the black-body problem and Poincar\'e's 1900 paper on the relativistic dynamics of charges and the electromagnetic field. These three papers could then be considered as part of Einstein's reflexions on the wave particle duality of light which he will pursue all his life long. However, from a strict historical perspective, what can only be ascertained up to now is that Einstein worked Planck's paper before March 1904, Poincar\'e's one before May 1906, and heard of the latter around 1903 since he was interested in Poincar\'e's 1902 \textit{Science and hypothesis} which contains chapters (translated in German in the \textit{Physikalische Zeitschrift}) with discussions on Lorentz theory, the reaction principle and on the need - or not - of an ether.

\end{document}